# A multiscale nonequilibrium model for melting of metal powder bed subjected to constant heat flux

Jin Wang[1, 2], Mo Yang[1], Yuwen Zhang[2, *]

[1]*School of Energy and Power Engineering, University of Shanghai for Science and Technology, Shanghai 200093, China*

[2]*Department of Mechanical and Aerospace Engineering, University of Missouri, Columbia, MO 65203, USA*

**ABSTRACT**

A multiscale numerical model based on nonequilibrium thermal effect for melting of metal powder bed subjected to constant heat flux is developed. The volume shrinkage due to density change is taken into account. The nonequilibrium model is discretized by an implicit finite difference method and solved numerically using an iterative tri-diagonal matrix algorithm. The evolutions of powder bed surface temperature and various interfacial locations as well as the melting temperature range during the melting process are investigated. The results show that liquid region, upper and lower parts of mushy zone are formed on the top of unsintered zone as the melting progresses. The duration of the preheating stage shortens and the melting rate accelerates as the initial porosity or initial temperature increases while particle size has much less effect on the melting process. The parametric study shows the melting temperature range of the powder bed widens with increasing initial porosity, decreasing initial temperature or increasing particle size.

**Keywords:** Direct Metal Laser Sintering, Melting, Multiscale Modeling, Powder, Particle

## 1. Introduction

Selective laser sintering (SLS) is a rapid prototyping and additive manufacturing technology that creates three-dimensional freeform and nearly full-density parts via layer-by-layer sintering or melting of metal powders induced by a directed laser beam [1-5]. For metal powder-based SLS process, the loose powder bed is heated and melted into a liquid pool by the incident laser beam, and then the liquid metal resolidifies as the laser beam moves away. Melting and resolidification process have significant effects on the quality and precision of final products [5].

Many investigations of metal powder melting and solidification have been carried out by numerical and experimental methods to understand mechanism of the SLS process. Fischer et al. [6] performed SLS experiments on titanium powder subjected to a nanosecond Nd: YAG laser with two different energy coupling mechanisms. Gusarov et al. [7] analyzed the effects of powder structural parameters on effective contact conductivity of SLS powder bed using a spherical coordinate numerical model. Konrad et al. investigated analytically melting and resolidification of metal powder subjected to nanosecond laser heating based on the powder bed model [8] and particle model [9], respectively. Kim and Sim [10] studied numerically thermal behavior and fluid flow during the transient and steady state laser melting of alloys using the enthalpy and apparent capacity methods. Pak and Plumb [11] developed both constant volume and constant porosity models for melting of a two-component powder bed.

---

[*] Corresponding author. Email: zhangyu@missouri.edu.

One major challenge of SLS is the balling phenomenon [12], in which a series of spherical grains with diameters approximately equal to that of the laser beam are formed because the melted powder particles stick each other due to surface tension. There are several technologies to overcome the balling phenomenon. one of which is direct metal laser sintering (DMLS) that can manufacture directly densified metal parts subjected to laser beam with lower scanning speed in the sintering surroundings filled with protective gas [13]. More investigations about the sintering behavior, microstructural features and heat transfer mechanism of various metal powders of DMLS were carried out [14-17]. The metal powder can be completely molten without balling phenomenon and the final densified parts do not need post-processing.

During the DMLS process, the melting accompanying with shrinkage due to significant density change makes it very complex. The effects of shrinkage on melting and resolidification during the DMLS process have been investigated using different models and methods. Zhang and Faghri [18] analytically solved melting of mixed powder bed with constant heat flux heating based on a one-dimensional model considering shrinkage. Xiao and Zhang [19] investigated numerically the shrinkage effects on temperature distribution and solid-liquid interface location of DMLS using the three-dimensional convection model. Childs et al. [20] studied density change and melting depths of single layer and multi-layer DMLS process based on experimental and three-dimensional numerical methods. Dai and Shaw [21] presented a three-dimensional thermomechanical finite element model to investigate the transient temperature, transient stresses, and residual stresses of powder bed. The numerical model encompassed the effects of density and temperature changes on the thermal conduction, radiation and natural convection. Kima et al. [22] investigated heat transfer and fluid flow of the molten pool in stationary gas tungsten arc welding considering shrinkage and convection induced by surface tension, buoyancy, and impinging plasma arc forces. Xiao and Zhang analytically solved the problems of partial melting [23] and complete melting [24] for powder bed with constant heat flux. The results showed that the shrinkage effects on the geometry and temperature distribution during the DMLS process were significant.

In contrast to melting and resolidification with well-defined melting point in conventional sintering, phase change for DMLS, from a macroscopic point of view, occur gradually within a temperature range of $(T_{sm}, T_{lm})$ due to the nonequilibrium thermal phenomenon [1], in which the temperature of the particle surface is much higher than that of the particle core because the energy heats up at the surface caused by the very high intensity of the laser beam. Therefore, during the DMLS process the mean temperature of the melting particle is within a range of temperature adjacent to the melting point, depending on the melting degree and temperature distribution of the particle. Gu and Shen [25] studied experimentally the development of WC-Co particulate reinforcing Cu matrix composite material using direct laser sintering and found the nonequilibrium effects played important roles on the melting process. Xiao and Zhang [23] developed an analytical model considering nonequilibrium thermal phenomenon in DMLS, but the temperature range was assumed to be known prior.

In order to understand the nonequilibrium thermal effects on melting of metal powder bed during the DMLS process, a multiscale nonequilibrium model for melting with constant heat flux considering shrinkage is developed in the present study. The model is based on nonequilibrium thermal effect that the melting temperature range of the powder bed is dependent on the degree of melting and temperature distribution of the particle. The effects of particle size, initial temperature and initial porosity of the powder bed on temperature distribution and melting geometry will be systematically investigated.

## 2. Problem statement and formulation

For the problem under consideration, the time scale of thermal diffusion within particles can be obtained by the time constant $t_p = \rho_p c_p V_p / h_f A_p$ [26], where $V_p$, $A_p$ are volume and surface area of the particle and heat transfer coefficient $h_f$ can be estimated by Nusselt number for heat conduction [26]; the time scale of the particle is in the order of $10^{-6}$s. On the other hand, the time scale of the laser sintering problem can be estimated based by the pulse duration ($10^{-7}$s) or the thermal penetration time of one particle depth [23] (about $10^{-7}$s). The time scale of the particle is larger than that of the macroscopic problem, so a multiscale model is required and established for the melting of metal powder bed during the DMLS process.

The multiscale model is consistent of two sub-models: one for powder bed level based on continuum assumption and the other one at the particle level to obtain the range of phase change temperature. Under the powder bed level model, the powder bed is treated as continuum that melts in a range of temperature; the temperature range depends on the degree of melting and temperature distribution in the particle. Meanwhile, the temperature distribution from the particle level model relates to the heat flux from the powder bed. Therefore, the multiscale nonequilibirium model is a coupled problem of the powder bed and the particle which can be solved iteratively.

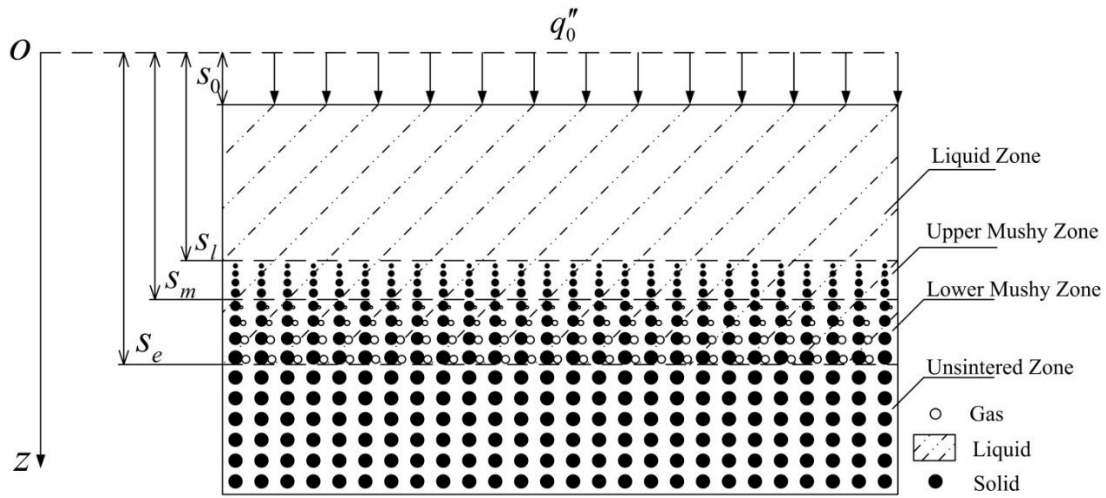

**Fig. 1** Physical model for melting of the metal powder bed with constant heat flux

### 2.1 Heat transfer in the powder bed

Figure 1 shows the physical model for metal powder bed with constant heat flux. The heating flux is absorbed by the randomly packed metal powder bed ($z \geq 0$) with an initial porosity of $\varepsilon_0$ and initial temperature of $T_i$, which is below the melting point of the metal powder material $T_m$. During the DMLS process, the powder particle undergoes melting has a mean temperature adjacent to the melting point due to nonequilibrium thermal effects of DMLS process [1]. Under constant heat flux heating, the surface temperature of the powder bed increases and then melting occurs when the powder bed surface temperature reaches to the lower limit of the range of melting temperature, $T_{sm}$, instead of the melting point $T_m$. A mushy zone where the solid and liquid coexist is formed as the temperature increases continuously, finally the liquid zone appears on the top of the mushy zone when the surface temperature increases to the upper limit $T_{lm}$. Consequently, there are three regions of the entire powder bed: liquid zone, mushy zone and unsintered zone.

During the melting process, the liquid permeates into the void of the unsintered zone as the interstitial gas between solid particles is driven out, and the solid core in the mushy zone moves downward due to shrinkage. The shrinkage is completed when the gas volume fraction of the mushy zone decreases to zero, so the mushy zone can be divided into two sub-regions: the lower part with shrinkage and the upper part without shrinkage. For the lower part, gas, liquid and solid coexist. It is assumed that the porosity defined as the void volume of gas and liquid relative to the total volume occupied by the solid core and void remains constant [24], the following volume fraction function for the lower part is satisfied.

$$\varepsilon_0 = \frac{V_l + V_g}{V_l + V_g + V_s} = \varphi_g + \varphi_l, \quad s_m(t) \leq z \leq s_e(t) \tag{1}$$

where $s_e(t)$, $s_m(t)$ are the locations of the mushy-solid interface and the mid-interface between the upper and lower mushy zones, respectively.

For the metal powder bed, the contribution of gas on mass can be neglected in mushy zone due to the lower density of gas compared with that of the particles. It is further assumed that the physical properties of metal particle for both solid and liquid phases including density, thermal conductivity and specific heat are the same. The solid mass fraction for the lower part of the mushy zone can be expressed as

$$f_s = \frac{\rho_P V_s}{\rho_P V_l + \rho_P V_s} = \frac{1-\varepsilon_0}{1-\varphi_g}, \quad s_m(t) \leq z \leq s_e(t) \tag{2}$$

During the melting process, the liquid volume fraction increases gradually from zero to one as the temperature increases from $T_{sm}$ to $T_{lm}$ in the whole mushy zone. Assuming linear liquid volume fraction function of local temperature by

$$\varphi_l = \begin{cases} 1 & T \geq T_{lm} \\ \dfrac{T-T_{sm}}{T_{lm}-T_{sm}} & T_{sm} < T < T_{lm} \\ 0 & T \leq T_{sm} \end{cases} \tag{3}$$

For the upper part of the mushy zone, there are only liquid and solid, so the volume fraction is the same as the mass fraction for both solid and liquid phases. And then the solid mass fraction for the upper part of the mushy zone can be expressed as

$$f_s = \begin{cases} 0 & T \geq T_{lm} \\ \dfrac{T_{lm} - T}{T_{lm} - T_{sm}} & T_{sm} < T < T_{lm} \\ 1 & T \leq T_{sm} \end{cases}, \quad s_l(t) \leq z < s_m(t) \tag{4}$$

where $s_l(t)$ is the locations of the liquid-mushy interface.

The entire process can be regarded as two sub-stages depending on whether melting occurs: preheating and melting stages. During the preheating stage, the temperature of the entire powder bed is below $T_{sm}$ and there is no melting until the surface temperature of the powder bed rises to $T_{sm}$. During the second stage, melting accompanied with shrinkage takes place.

### 2.1.1 Preheating stage

During the preheating stage, the temperature of the powder bed increases as the surface absorbs the heat. The heat transfer of this stage can be described as pure conduction so the energy equation can be expressed as

$$(1-\varepsilon_0)\rho_p c_p \frac{\partial T}{\partial t} = k_{eff} \frac{\partial^2 T}{\partial z^2}, \quad z > 0, 0 < t \leq t_m \tag{5}$$

where $t_m$ is the time at which the temperature of the bed surface reaches to $T_{sm}$ and the melting stage begins, and $k_{eff}$ is the effective thermal conductivity of the unsintered powder bed. It can be calculated using the empirical correlation obtained by Hadley [27] for the randomly packed powder bed in the present study.

$$k_{eff} = k_g(1-\alpha_0)\frac{\varepsilon_0 f_0 + (k_p/k_g)(1-\varepsilon_0 f_0)}{1-\varepsilon_0(1-f_0)+(k_p/k_g)\varepsilon_0(1-f_0)} + \alpha_0 \frac{2(k_p/k_g)^2(1-\varepsilon_0)+(1+2\varepsilon_0)(k_p/k_g)}{(1-\varepsilon_0)+(k_p/k_g)(2+\varepsilon_0)} \tag{6}$$

where

$$f_0 = 0.8 + 0.1\varepsilon_0 \tag{7}$$

$$\log \alpha_0 = \begin{cases} -4.898\varepsilon_0 & 0 \leq \varepsilon_0 \leq 0.0827 \\ -0.405 - 3.154(\varepsilon_0 - 0.0827) & 0.0827 < \varepsilon_0 \leq 0.298 \\ -1.084 - 6.778(\varepsilon_0 - 0.298) & 0.298 < \varepsilon_0 \leq 0.58 \end{cases} \tag{8}$$

For the preheating stage, the initial and boundary conditions can be given by

$$T = T_i, \quad z \geq 0, t = 0 \tag{9}$$

$$-k_{eff}\frac{\partial T}{\partial z} = q_0'', \quad z = 0, 0 < t \leq t_m \tag{10}$$

$$T = T_i, \ z \to +\infty, 0 < t \leq t_m \tag{11}$$

where $q_0''$ is the intensity of the constant heat flux.

*2.1.2 Melting stage*

During the melting stage, the powder bed is divided into four regions from the heating surface to the bottom: the liquid zone, the upper and lower mushy zone, and the unsintered region as the analyzed above.

For the liquid zone, the energy equation can be expressed as

$$\rho_p c_p \frac{\partial T}{\partial t} + \rho_p c_p w_0 \frac{\partial T}{\partial z} = k_p \frac{\partial^2 T}{\partial z^2}, \ s_0(t) < z < s_l(t), t > t_m \tag{12}$$

where $s_0(t)$ is the locations of the heating surface, and $w_0$ is the velocity due to shrinkage, which can be obtained by

$$w_0 = \frac{ds_0(t)}{dt} \tag{13}$$

The boundary condition is expressed as

$$-k_p \frac{\partial T}{\partial z} = q_0'', \ z = s_0(t), t > t_m \tag{14}$$

For the upper part of the mushy zone, the energy equation can be given as

$$\rho_p c_p \frac{\partial T}{\partial t} + \rho_p c_p w_0 \frac{\partial T}{\partial z} = k_p \frac{\partial^2 T}{\partial z^2} + \rho_p h_{sl} \frac{\partial f_s}{\partial t}, \ s_l(t) < z < s_m(t), t > t_m \tag{15}$$

which is subjected to the following boundary conditions:

$$\left(k_p \frac{\partial T}{\partial z}\right)_{z=s_l^+(t)} = \left(k_p \frac{\partial T}{\partial z}\right)_{z=s_l^-(t)}, \ z = s_l(t), t > t_m \tag{16}$$

$$T = T_{lm}, \ z = s_l(t), t > t_m \tag{17}$$

For the lower part of the mushy zone, the energy and continuity equations can be expressed as

$$\frac{\partial\left((1-\varphi_g)\rho_p c_p T\right)}{\partial t} + \frac{\partial\left((1-\varphi_g)\rho_p c_p wT\right)}{\partial z} = \frac{\partial}{\partial z}\left(k_m \frac{\partial T}{\partial z}\right) + (1-\varphi_g)\rho_p h_{sl} \frac{\partial f_s}{\partial t},$$
$$s_m(t) < z < s_e(t), t > t_m \tag{18}$$

$$\frac{\partial(1-\varphi_g)}{\partial t} + \frac{\partial\left((1-\varphi_g)w\right)}{\partial z} = 0, \ s_m(t) < z < s_e(t), t > t_m \tag{19}$$

where $k_m$, the thermal conductivity in the lower mushy zone, is much higher than that before melting because the contact area increase as melting progresses. It is assumed as a linear function of local porosity, it can be analytically expressed as

$$k_m = (1-\varphi_g)k_p \tag{20}$$

Substituting Eqs. (2) and (20) into Eqs. (18) and (19), the governing equations can be rewritten as

$$\frac{\partial}{\partial t}\left(\rho_p c_p \frac{T}{f_s}\right) + \frac{\partial}{\partial z}\left(\rho_p c_p w \frac{T}{f_s}\right) = \frac{\partial}{\partial z}\left(\frac{k_p}{f_s}\frac{\partial T}{\partial z}\right) + \frac{\rho_p h_{sl}}{f_s}\frac{\partial f_s}{\partial t}, \ s_m(t) < z < s_e(t), t > t_m \tag{21}$$

$$\frac{\partial}{\partial t}\left(\frac{1}{f_s}\right) + \frac{\partial}{\partial z}\left(\frac{w}{f_s}\right) = 0, \ s_m(t) < z < s_e(t), t > t_m \tag{22}$$

and the boundary conditions of the lower mushy zone are

$$\left(k_m \frac{\partial T}{\partial z}\right)_{z=s_m^+(t)} = \left(k_p \frac{\partial T}{\partial z}\right)_{z=s_m^-(t)}, \ z = s_m(t), t > t_m \tag{23}$$

At the mid-interface $s_m(t)$, the liquid volume fraction is $\varepsilon_0$, so according to Eq. (4) the temperature at the mid-interface can be obtained by

$$T = T_{sm} + \varepsilon_0 (T_{lm} - T_{sm}), \ z = s_m(t), t > t_m \tag{24}$$

For the unsintered solid zone, the energy equation is the same as that of the preheating stage (Eq. (5)) with the following boundary conditions:

$$\left(k_{eff}\frac{\partial T}{\partial z}\right)_{z=s_e^+(t)} = \left(k_m \frac{\partial T}{\partial z}\right)_{z=s_e^-(t)}, \ z > s_e(t), t > t_m \tag{25}$$

$$T = T_{sm}, \ z = s_e(t), t > t_m \tag{26}$$

$$T = T_i, \ z \to +\infty, t > t_m \tag{27}$$

### 2.2 Heat transfer in the particle

The melting temperature range $(T_{sm}, T_{lm})$ in the powder bed level model is related to the melting degree and temperature distribution of the metal particle; it can be obtained from the particle level model. Figure 2 shows the physical model for a melting metal particle with a radius of $r_p$. The particle with uniform initial temperature $T_i$ absorbs the heat at its surface. It is assumed that the sphere shape is not change during the melting because of the rapid melting velocity and the quite smaller size of the particle compared with that of the powder bed. The governing equations for the particle based on the coordinate system setting the origin at the particle surface are written as

$$\frac{\partial T_{pl}}{\partial t} = \frac{\alpha_p}{(r_p - x)^2}\frac{\partial}{\partial x}\left((r_p - x)^2 \frac{\partial T_{pl}}{\partial x}\right), \ 0 < x < x_m(z,t) \tag{28}$$

$$\frac{\partial T_{ps}}{\partial t} = \frac{\alpha_p}{(r_p - x)^2}\frac{\partial}{\partial x}\left((r_p - x)^2 \frac{\partial T_{ps}}{\partial x}\right), \ x_m(z,t) < x < r_p \tag{29}$$

which are subject to the following initial and boundary conditions::

$$-k_p \frac{\partial T_{pl}}{\partial x} = q''(z,t), \ x = 0 \tag{30}$$

$$-k_p \frac{\partial T_{pl}}{\partial x} + k_p \frac{\partial T_{ps}}{\partial x} = \rho_p h_{sl} \frac{\partial x_m(z,t)}{\partial t}, \ x = x_m(z,t) \tag{31}$$

$$\frac{\partial T_{ps}}{\partial x} = 0, \ x = r_p \tag{32}$$

$$T_{ps} = T_i, \ 0 \le x \le r_p, t = 0 \tag{33}$$

where $q''(z,t)$ is the heat flux to the particle. It is related to the time and the particle location in the powder bed, and can be calculated using the equivalent heat flux method based on the energy conservation by

$$\left((1-\varphi_g)\rho_p c_p \frac{\partial T}{\partial t} + (1-\varphi_g)\rho_p h_{sl} \frac{\partial f_l}{\partial t}\right)\Delta V = 4\pi r_p^2 n_p q''(z,t) \tag{34}$$

where $\Delta V$, $n_p$ are the volume element of the powder bed and the particle number in the volume element, respectively. According to the definition of the volume fraction, the relation $\Delta V$ and $n_p$ can be obtained by

$$\varphi_g = \frac{V_g}{V_g + V_l + V_s} = \frac{\Delta V - n_p \cdot 4\pi r_p^3/3}{\Delta V} \tag{35}$$

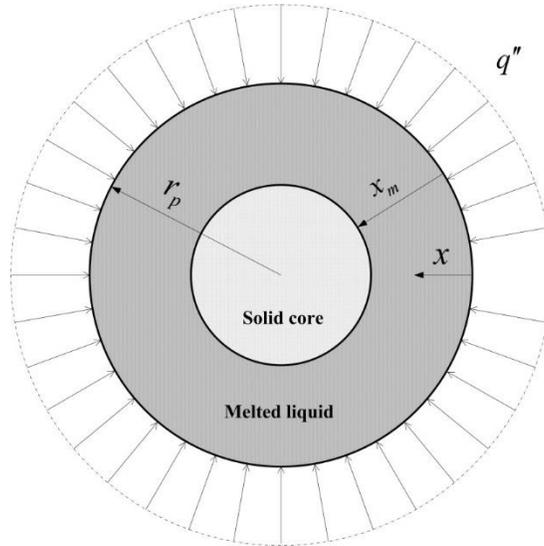

**Fig. 2** The physical model for a melting metal particle

Substituting Eq. (35) into Eq. (34) the heat flux of the particle become

$$q''(z,t) = \frac{r_p}{3}\left(\rho_p c_p \frac{\partial T}{\partial t} + \rho_p h_{sl} \frac{\partial f_l}{\partial t}\right) \tag{36}$$

As the heat flux is absorbed, the temperature of the particle surface increases. And the melting of the particle occurs when the surface temperature reaches to $T_m$. At the moment $t = t_{sm}$ the average temperature of the particle is $T_{sm}$ and it can be calculated by

$$T_{sm} = \frac{\left(\int_0^{r_p} 4c_p \pi (r_p - x)^2 T_{ps} dx\right)_{t=t_{sm}}}{4c_p \pi r_p^3/3} \tag{37}$$

When the temperature of the center of the particle sphere rises to $T_m$ as melting processes, the melting of the particle is complete and the average temperature of $T_{lm}$ is given by

$$T_{lm} = \frac{\left(\int_0^{r_p} 4c_p \pi (r_p - x)^2 T_{pl} dx\right)_{t=t_{lm}}}{4c_p \pi r_p^3/3} \tag{38}$$

## 2.3 Dimensionless equations

The above multiscale nonequilibrium model can be non-dimensionalized using the following dimensionless variables.

$$\theta = \frac{c_p(T-T_i)}{h_{sl}}, \quad \tau = \frac{\alpha_p t}{(\alpha_p \rho_p h_{sl}/q_0'')^2}, \quad K_{eff} = \frac{k_{eff}}{k_p(1-\varepsilon_0)}, \quad Z = \frac{z}{\alpha_p \rho_p h_{sl}/q_0''}, \quad W = \frac{\rho_p h_{sl} w}{q_0''},$$

$$W_0 = \frac{\rho_p h_{sl} w_0}{q_0''}, \quad \theta_{lm} = \frac{c_p(T_{lm}-T_i)}{h_{sl}}, \quad \theta_{sm} = \frac{c_p(T_{sm}-T_i)}{h_{sl}}, \quad \theta_m = \frac{c_p(T_m-T_i)}{h_{sl}}, \quad \theta_{pl} = \frac{c_p(T_{pl}-T_i)}{h_{sl}},$$

$$\theta_{ps} = \frac{c_p(T_{ps}-T_i)}{h_{sl}}, \quad X = \frac{x}{r_p}, \quad D_p = \frac{r_p}{\alpha_p \rho_p h_{sl}/q_0''}$$

The governing equations and conditions can be obtained in dimensionless form as follows:

*2.3.1 Preheating stage*

$$\frac{\partial \theta}{\partial \tau} = K_{eff} \frac{\partial^2 \theta}{\partial Z^2}, \quad Z > 0, 0 < \tau \leq \tau_m \tag{39}$$

$$\theta = 0, \quad Z \geq 0, \tau = 0 \tag{40}$$

$$\frac{\partial \theta}{\partial Z} = \frac{1}{K_{eff}(\varepsilon_0 - 1)}, \quad Z = 0, 0 < \tau \leq \tau_m \tag{41}$$

$$\theta = 0, \quad Z \to +\infty, 0 < \tau \leq \tau_m \tag{42}$$

*2.3.2 Liquid zone of melting stage*

$$\frac{\partial \theta}{\partial \tau} + W_0 \frac{\partial \theta}{\partial Z} = \frac{\partial^2 \theta}{\partial Z^2}, \ S_0(\tau) < Z < S_l(\tau), \tau > \tau_m \tag{43}$$

$$\frac{\partial \theta}{\partial Z} = -1, \ Z = S_0(\tau), \tau > \tau_m \tag{44}$$

*2.3.3 Upper mushy zone of melting stage*

$$\frac{\partial \theta}{\partial \tau} + W_0 \frac{\partial \theta}{\partial Z} = \frac{\partial^2 \theta}{\partial Z^2} + \frac{\partial f_s}{\partial \tau}, \ S_l(\tau) < Z < S_m(\tau), \tau > \tau_m \tag{45}$$

$$\left(\frac{\partial \theta}{\partial Z}\right)_{Z=S_l^+(\tau)} = \left(\frac{\partial \theta}{\partial Z}\right)_{Z=S_l^-(\tau)}, \ Z = S_l(\tau), \tau > \tau_m \tag{46}$$

$$\theta = \theta_{lm}, \ Z = S_l(\tau), \tau > \tau_m \tag{47}$$

*2.3.4 Lower mushy zone of melting stage*

$$\frac{\partial}{\partial \tau}\left(\frac{\theta}{f_s}\right) + \frac{\partial}{\partial Z}\left(W \frac{\theta}{f_s}\right) = \frac{\partial}{\partial Z}\left(\frac{1}{f_s}\frac{\partial \theta}{\partial Z}\right) + \frac{1}{f_s}\frac{\partial f_s}{\partial \tau}, \ S_m(\tau) < Z < S_e(\tau), \tau > \tau_m \tag{48}$$

$$\frac{\partial}{\partial \tau}\left(\frac{1}{f_s}\right) + \frac{\partial}{\partial Z}\left(\frac{W}{f_s}\right) = 0, \ S_m(\tau) < Z < S_e(\tau), \tau > \tau_m \tag{49}$$

$$\left(\frac{1-\varepsilon_0}{f_s}\frac{\partial \theta}{\partial Z}\right)_{Z=S_m^+(\tau)} = \left(\frac{\partial \theta}{\partial Z}\right)_{Z=S_m^-(\tau)}, \ Z = S_m(\tau), \tau > \tau_m \tag{50}$$

$$\theta = \theta_{sm} + \varepsilon_0(\theta_{lm} - \theta_{sm}), \ Z = S_m(\tau), \tau > \tau_m \tag{51}$$

*2.3.5 Unsintered zone of melting stage*

$$\frac{\partial \theta}{\partial \tau} = K_{eff} \frac{\partial^2 \theta}{\partial Z^2}, \ Z > S_e(\tau), \tau > \tau_m \tag{52}$$

$$\left(f_s K_{eff} \frac{\partial \theta}{\partial Z}\right)_{Z=S_e^+(\tau)} = \left(\frac{\partial \theta}{\partial Z}\right)_{Z=S_e^-(\tau)}, \ Z = S_e(\tau), \tau > \tau_m \tag{53}$$

$$\theta = \theta_{sm}, \ Z = S_e(\tau), \tau > \tau_m \tag{54}$$

$$\theta = 0, \ Z \to +\infty, \tau > \tau_m \tag{55}$$

*2.3.6 Melting of the particle*

$$\frac{\partial \theta_{pl}}{\partial \tau} = \frac{1}{D_P^2(1-X)^2}\frac{\partial}{\partial X}\left((1-X)^2 \frac{\partial \theta_{pl}}{\partial X}\right), \ 0 < X < X_m(Z, \tau) \tag{56}$$

$$\frac{\partial \theta_{ps}}{\partial \tau} = \frac{1}{D_P^2(1-X)^2}\frac{\partial}{\partial X}\left((1-X)^2 \frac{\partial \theta_{ps}}{\partial X}\right), \ X_m(Z, \tau) < X < 1 \tag{57}$$

$$-\frac{\partial \theta_{pl}}{\partial X} = \frac{D_p^2}{3}\left(\frac{\partial \theta}{\partial \tau} - \frac{\partial f_s}{\partial \tau}\right), \quad X = 0 \tag{58}$$

$$-\frac{\partial \theta_{pl}}{\partial X} + \frac{\partial \theta_{ps}}{\partial X} = D_p^2 \frac{\partial X_m(Z,\tau)}{\partial \tau}, \quad X = X_m(Z,\tau) \tag{59}$$

$$\frac{\partial \theta_{ps}}{\partial X} = 0, \quad X = 1 \tag{60}$$

$$\theta_{ps} = 0, \quad 0 \leq X \leq 1, \tau = 0 \tag{61}$$

$$\theta_{sm} = 3\left(\int_0^1 (1-X)^2 \theta_{ps} dX\right)_{\tau=\tau_{sm}} \tag{62}$$

$$\theta_{lm} = 3\left(\int_0^1 (1-X)^2 \theta_{pl} dX\right)_{\tau=\tau_{lm}} \tag{63}$$

## 3. Numerical method

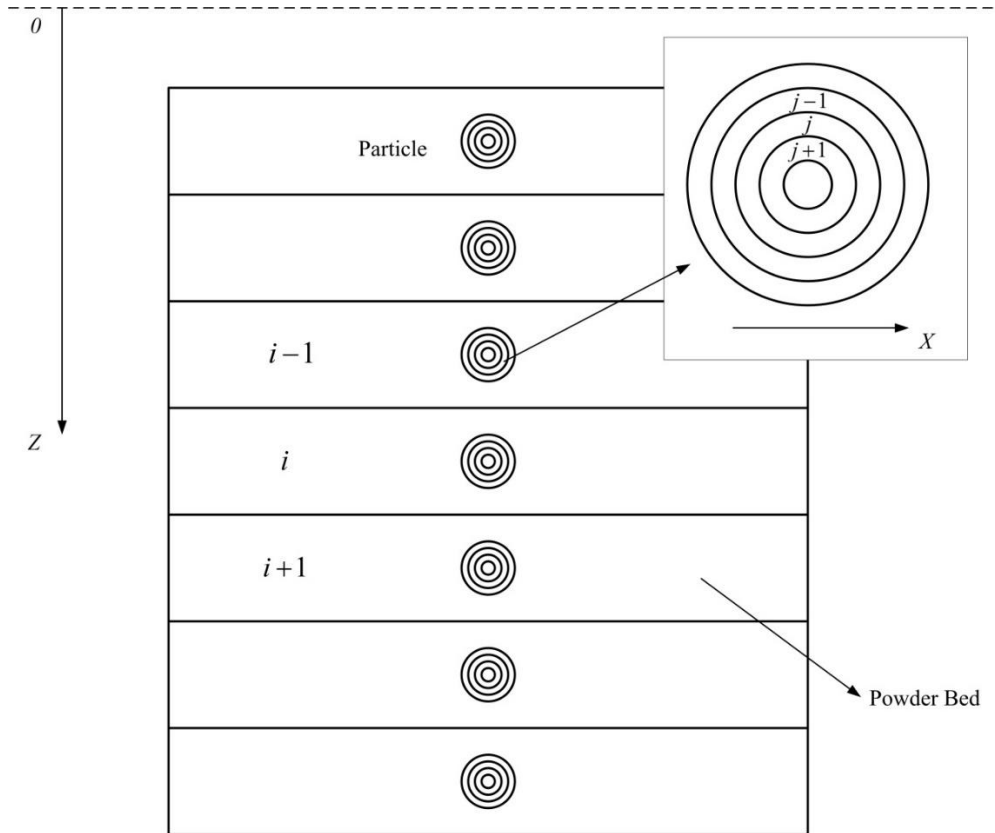

**Fig. 3** Gird system for the multiscale melting model

The melting process under consideration is a multiscale phase change problem which can be discretized by the implicit finite difference method [28] in two types of grid systems: powder bed grids and particle grids as Fig. 3. The discretized equations are solved by the tri-diagonal matrix algorithm (TDMA) method. Since the melting temperature range $(T_{sm}, T_{lm})$ depends on the degree of melting of the metal particle and the particle location, it is necessary to solve the temperature distribution of the particle first. Meanwhile, the particle temperature distribution relates to the heat flux of the local particle from the powder bed. Therefore, the iteration method is employed to solve the coupled problems of the powder bed and the particle. And the iterating criterion on one time step is established as follow,

$$\sum_i \left| \phi_i^n - \phi_i^{n-1} \right| < 10^{-6} \tag{64}$$

where $\phi$ is the generic variable that can be $\theta$ or $S$, and the superscript $n$ indicates the iteration number in one time step. The subscript sequence $i$ represents the grid node of the powder bed. A block-off technique [29] is used to simulate the empty zone of the computational domain created by shrinkage of the powder bed due to shrinkage.

Effects of the gird size and time step on the melting process must be studied before simulations are carried out. Three different gird systems, namely, $1200 \times 200$ (in the $Z$ direction of the powder bed and the $X$ direction of the particle, respectively), $1000 \times 100$, and $800 \times 50$ are employed to simulate the melting process. Considering simulated accuracy and CPU time in the range of variables, the grid of $1000 \times 100$ is sufficiently fine to ensure the grid independent solution. Meanwhile, the time step independence has been examined and the dimensionless time step of 0.0002 is applied. The initial porosity of the powder bed, initial temperature and particle size play significant roles for the multiscale model during DMLS. The evolution of the surface temperature of the powder bed, as well as the location of the mushy-solid, midi-interface and liquid-mushy interface, and the top surface of the powder bed will be evaluated.

## 4. Results and discussions

Table 1 Thermophysical properties of Nickel Braze [31]

| Property | Unit | Value |
| --- | --- | --- |
| Density, $\rho_P$ | $kg/m^3$ | 8257 |
| Thermal conductivity, $k_P$ | $W/(m \cdot K)$ | 14.65 |
| Specific heat, $c_P$ | $J/(kg \cdot K)$ | 462.6 |
| Melting point, $T_m$ | $K$ | 1258 |
| Fusion latent heat, $h_{sl}$ | $kJ/kg$ | 377.4 |

To demonstrate the validity of the simulation code for the multiscale nonequilibrium thermal model of melting with constant heat flux during DMLS, the developed numerical program is first applied to simulate (1) laser melting process of nickel braze powder bed with constant heat flux [23], and (2) powder particle melting process subjected to nanosecond laser heating [9]. For the powder bed melting process, the spherical powder particles with uniform initial temperature $T_i = 293K$ and particle size arrange as face-centered cubic corresponding to the initial porosity of 0.264 [30]. The thermophysical properties of nickel braze used in the simulations are shown in Table 1 [31]. The dimensionless temperature range here is regarded as constant is chosen to be 0.034. The dimensionless thermal conductivity of gas in the powder bed interstitial is $K_g = 3.7 \times 10^{-4}$. Figure 4 shows comparison of surface temperature evolutions of the powder bed during the melting process between present model and Xiao and Zhang [23], which are in good agreement with each other.

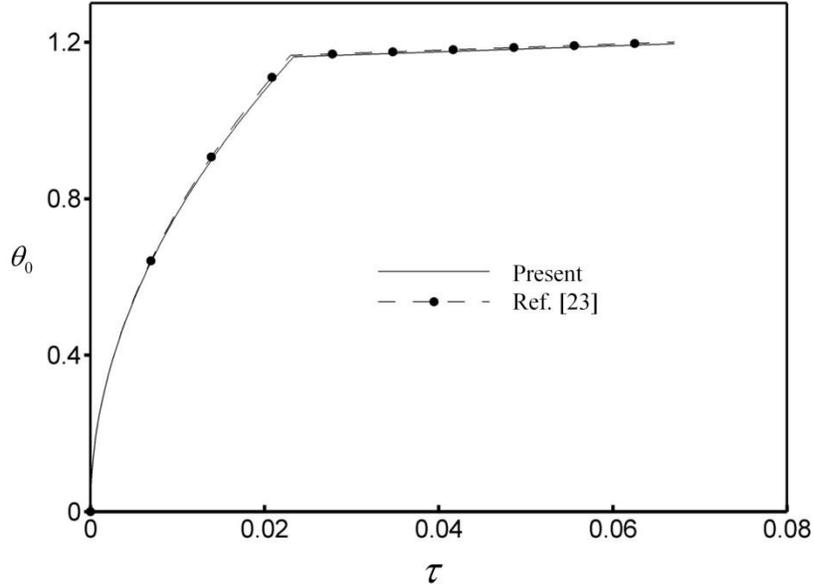

**Fig. 4** Comparison of Surface temperature of the powder bed during the melting process

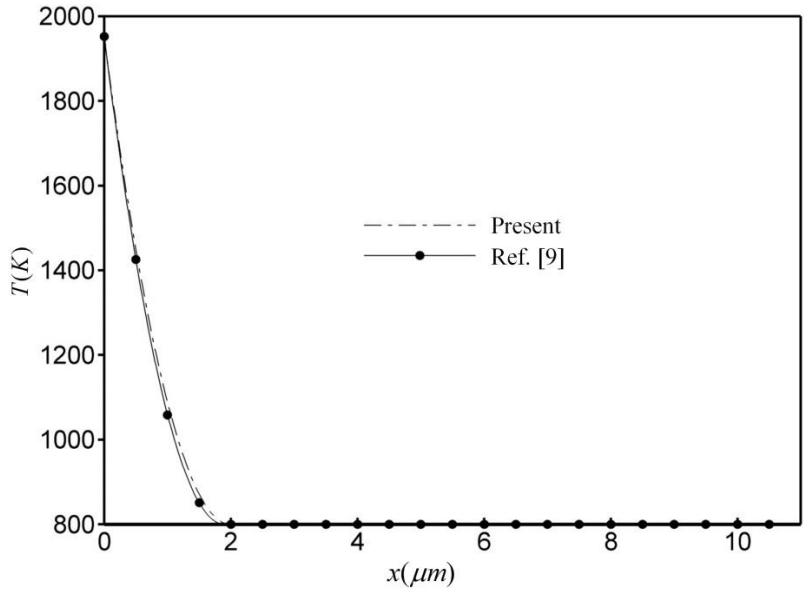

**Fig. 5** Comparison of temperature profile within the powder particle at $\tau = \tau_m$

For the particle melting process [9], the particle with uniform initial temperature of $T_i = 800K$ and radius of $r_p = 11\mu m$ is heated by nanosecond laser beam and melts at a well-defined melting point of $T_m = 1952K$. Figure 5 shows the temperature distributions of the powder particle at the beginning of the melting stage $\tau = \tau_m$ obtained by different methods. It can be seen that the results are in good agreement with each other. Moreover, the temperature of the particle surface is much higher than the center, i.e., the mean temperature of the particle is lower than the melting point while the melting occurs. Therefore, it is necessary to take the nonequilibrium thermal effect into account for the simulation of the powder bed melting.

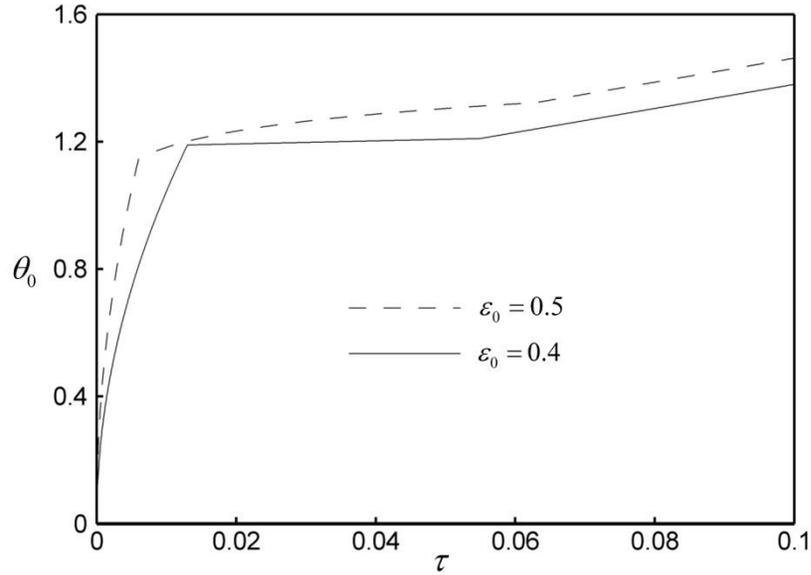

**Fig. 6** Evolution of surface temperature of powder bed for different initial porosities ($\theta_m = 1.2$, $D_p = 0.08$)

Numerical simulation is then carried out for melting of nickel braze powder bed subjected to constant heat flux based on the multiscale nonequilibrium model. Figure 6 presents the evolution of surface temperature of the powder bed for different initial porosities during the melting process. The temperature increases rapidly during the preheating stage due to the low effective thermal conductivity of the unsintered powder bed. At the beginning of the melting stage, the surface temperature increase slowly due to the fusion latent heat and sharply increased thermal conductivity as melting progresses. Once the liquid region on the surface of the powder bed is formed, the increasing rate of the temperature becomes faster. Meanwhile, it can be observed that the duration of preheating stage shortens with increasing initial porosity owing to the lower effective thermal conductivity for higher porosity. The melting temperature range at the powder bed surface become larger and the duration of the liquid region formation on the surface gets longer as initial porosity increases. The temperature difference between the particle surface and core larger and thus the nonequilibrium effect is greater for high-porosity powder bed.

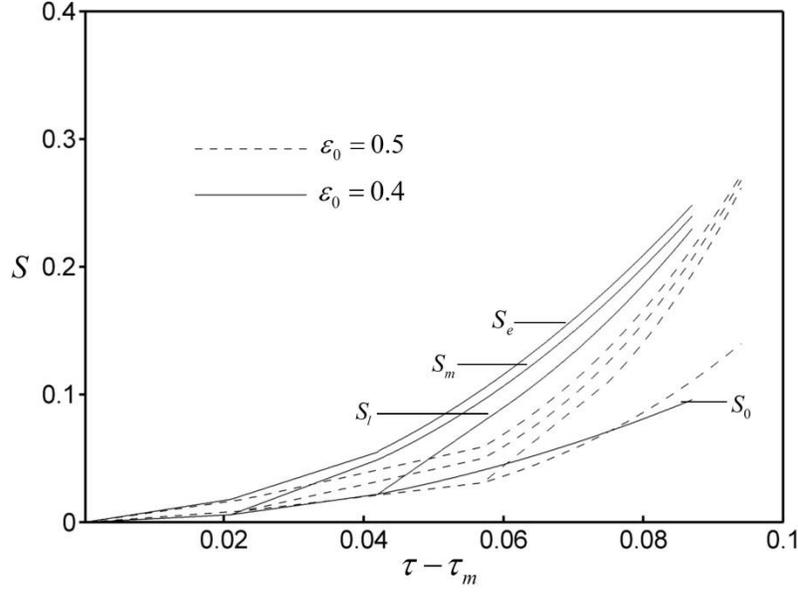

**Fig. 7** Interfacial locations for different initial porosities during the melting stage ($\theta_m = 1.2$, $D_p = 0.08$)

Figure 7 illustrates the evolution of various interfacial locations for different initial porosities during the melting stage corresponding to the same conditions as Fig. 6. At the beginning of the melting stage, the downward motion of the mushy-solid interface $S_e$ is faster than that of the heating surface $S_0$ and the lower part of the mushy zone is formed. Once the upper part of the mushy zone is formed as melting progresses, the downward moving speed of the mushy-solid interface becomes faster owing to higher thermal conductivity in the upper part. However, the speed of the mid-interface $S_m$ is faster than that of the mushy-solid interface and the lower part of the mushy zone narrows. After the liquid region is formed, the liquid-mushy interface $S_l$ moves more rapidly than the mid-interface and mushy-solid interface; thus, the whole mushy zone becomes narrower as the liquid region does not absorb the latent heat. In addition, although the greater nonequilibrium effect makes the duration of the liquid region formation longer for the case with higher porosity as shown in Fig. 6, the melting accelerates as initial porosity increases after the liquid zone is formed as shown in Fig. 7 due to lower thermal conductivity and density for high-high-porosity powder bed.

Figure 8 shows the effect of initial porosity on the melting temperature range at the powder bed surface corresponding to the same conditions as Fig. 6. Note that the upper limit of the melting temperature range $\theta_{lm}$ increases and the lower limit $\theta_{sm}$ decreases with increasing initial porosity due to the greater nonequilibrium effect for higher porosity as mentioned above. As initial porosity increases, the thermal conductivity of the loose powder bed decreases and more heat is available to heat and melt the powder particles at the particle surface;; the temperature difference between the particle surface and core increases and the melting temperature range ($\theta_{sm}, \theta_{lm}$) widens with increasing temperature difference. Another observation is that the change of the upper limit is shaper than the lower limit owing to the fusion latent heat as melting occurs.

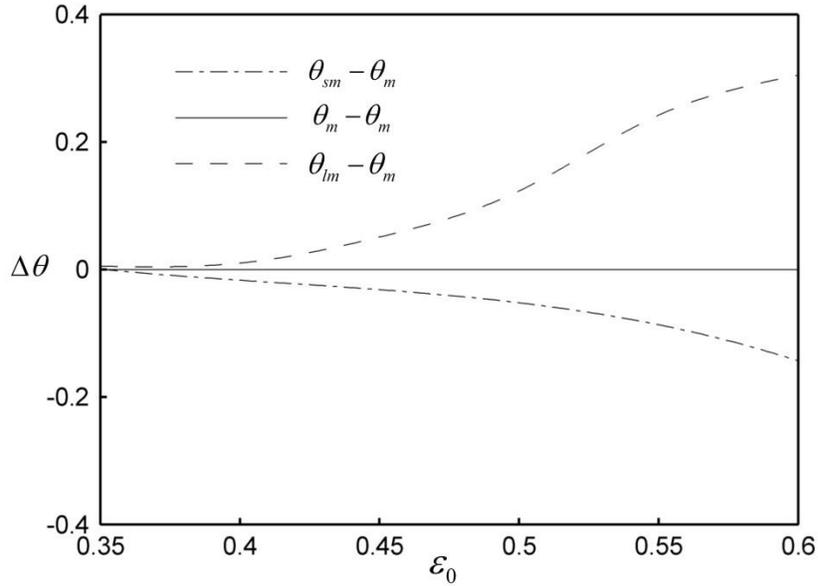

**Fig. 8** Effect of initial porosity on the melting temperature range at the powder bed surface ($\theta_m = 1.2$, $D_p = 0.08$)

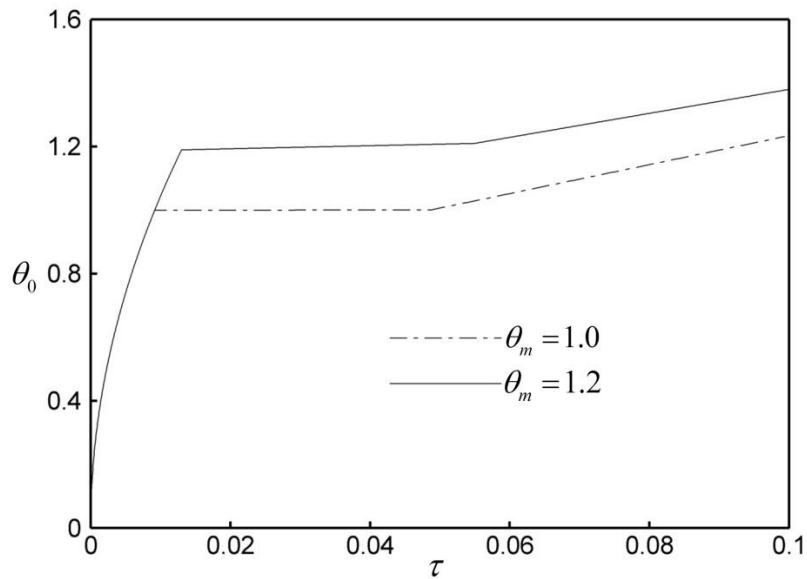

**Fig. 9** Evolution of surface temperature of powder bed for different initial temperatures ($\varepsilon_0 = 0.4$, $D_p = 0.08$)

Figure 9 shows the evolution of surface temperature of the powder bed for different initial temperatures during the melting process. As can be seen, the duration of preheating stage shortens as the dimensionless temperature $\theta_m$ decreases since the higher initial temperature corresponds to lower $\theta_m$. Figure 10 shows the evolution of various interfacial locations for different initial temperature corresponding to the same conditions as Fig. 9. It can be observed that the duration of the liquid region formation shortens and during this duration the four interfaces move a little faster with decreasing $\theta_m$ (or increasing initial temperature). After the liquid region is formed, the motion speeds of the four interfaces become much faster as initial temperature increases. Figure 11 shows the effect of initial temperature on the melting temperature range at powder bed surface corresponding to the same conditions as Fig. 9. The melting temperature range increases slightly as the dimensionless temperature $\theta_m$ increases, because the lower initial temperature for larger $\theta_m$ makes the temperature difference between the particle surface and core larger during the melting process and the nonequilibrium effect is greater as initial temperature decreases. However, compared with Fig. 8, the change trend of the temperature range with decreasing initial temperature is much less than the trend caused by increasing porosity.

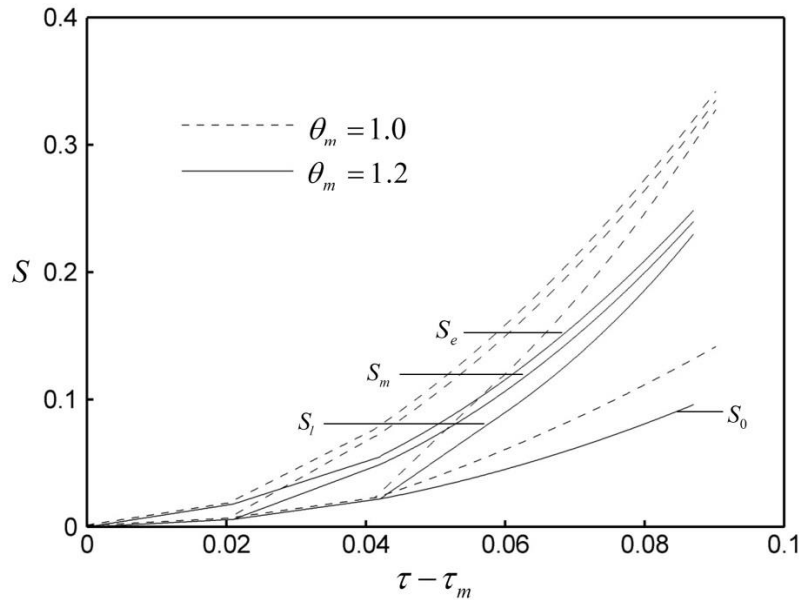

**Fig. 10** Interfacial locations for different initial temperatures during the melting stage ($\varepsilon_0 = 0.4$, $D_p = 0.08$)

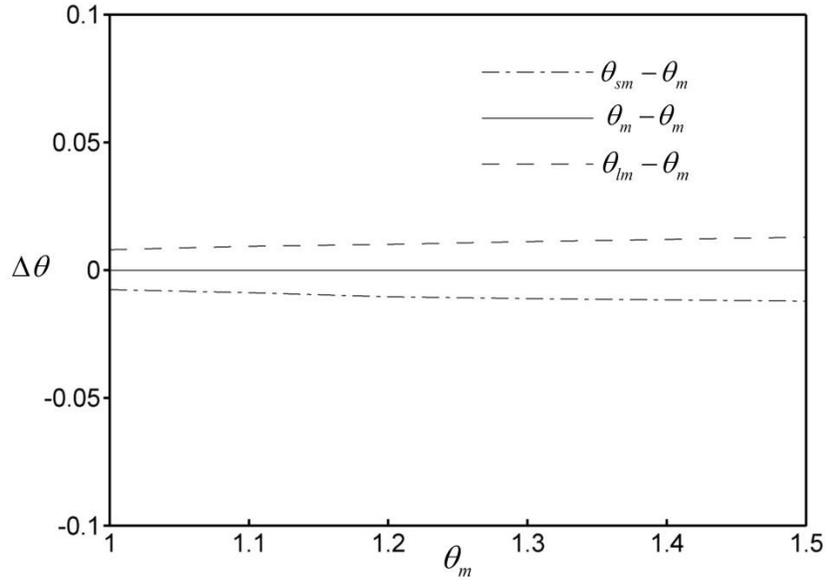

**Fig. 11** Effect of initial temperature on the melting temperature range at the powder bed surface ($\varepsilon_0 = 0.4$, $D_p = 0.08$)

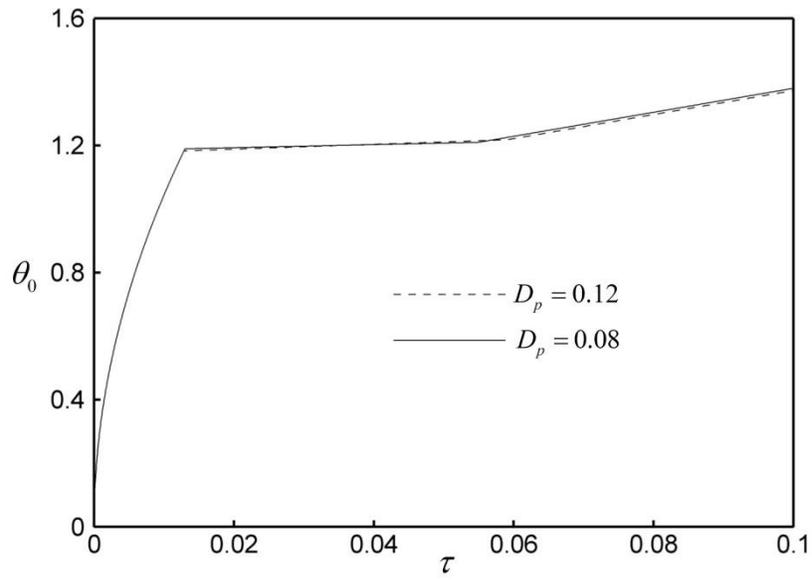

**Fig. 12** Evolution of surface temperature of powder bed for different particle sizes ($\varepsilon_0 = 0.4$, $\theta_m = 1.2$)

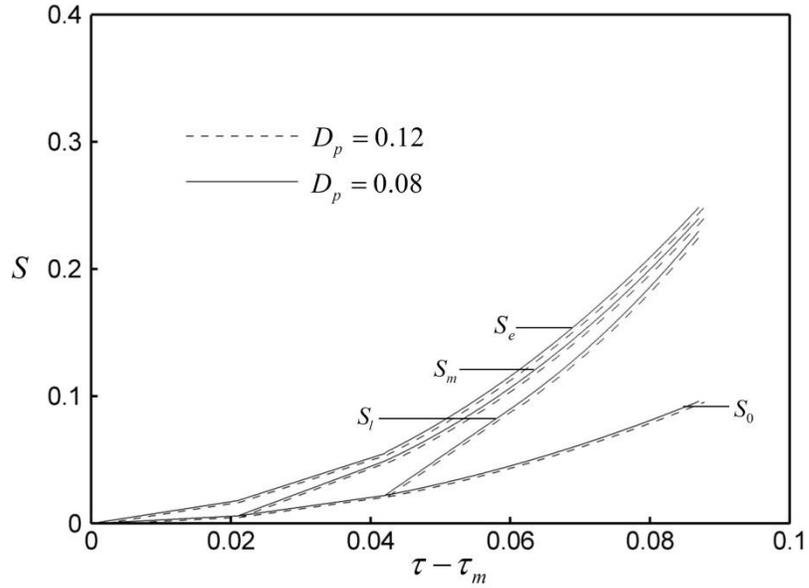

**Fig. 13** Interfacial locations for different particle sizes during the melting stage ($\varepsilon_0 = 0.4$, $\theta_m = 1.2$)

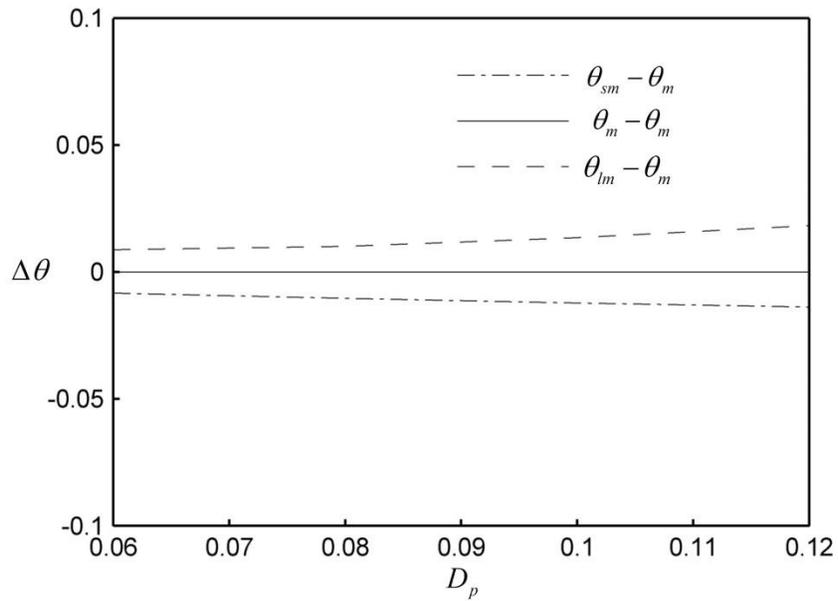

**Fig. 14** Effect of particle size on the melting temperature range at the powder bed surface ($\varepsilon_0 = 0.4$, $\theta_m = 1.2$)

Figure 12 shows the evolution of surface temperature of the powder bed for different particle size during the melting process. The effect of particle size on the surface temperature evolution is smaller than that of initial porosity as shown in Fig. 6 or initial temperature as shown in Fig. 9. Figure 13 presents the evolution of various interfacial locations for different particle sizes corresponding to the same conditions as Fig. 12. As particle size $D_p$ increases, the duration of the liquid region formation becomes slightly longer. And the moving speeds of the interfaces for the larger particle are slightly slower than those for the small particle. In other words, the increase of the particle size does not have significant influence on the motion of the interfaces during the melting stage. Figure 14 shows the effect of particle size on the melting temperature range at powder bed surface corresponding to the same conditions as Fig. 12. It can be seen that the melting temperature range increases slightly with increasing particle size because the lager particle makes the temperature difference between the particle surface and core larger.

**Conclusions**

A multiscale nonequilibrium thermal model for melting of metal powder bed subjected to constant heat flux with volume shrinkage has been presented. The model has been validated by comparing surface temperature evolution of powder bed and temperature distribution of particle with analytical solutions. The evolutions of powder bed surface temperature and various interfacial locations as well as the melting temperature range of the powder bed surface during the melting process have been investigated. The results show that liquid region, upper and lower parts of mushy zone are formed on the unsintered zone during the melting process. The preheating duration shortens and the melting rate after the liquid region is formed accelerates with increasing initial porosity or increasing initial temperature while particle size has much less effect on the melting process. The melting temperature range of the powder bed surface widens with increasing initial porosity, decreasing initial temperature or increasing particle size, but the effect of initial porosity on the melting temperature range is more significant than the effects of the initial temperature and particle size.

**Acknowledgment**

This work is supported by Chinese National Natural Science Foundations under Grants 51129602 and 51076105, Innovation Program of Shanghai Municipal Education Commission 14ZZ134, Innovation Fund Project for Graduate Student of Shanghai JWCXSL1301, SQI Commonweal Project No. 2012-12, and US National Science Foundation under Grant No. 1066917.

**Nomenclature**

| | |
|---|---|
| $w$ | liquid velocity, $m/s$ |
| $W$ | liquid dimensionless velocity |
| $T_m$ | melting temperature, $K$ |
| $T_i$ | initial temperature, $K$ |
| $q_0''$ | constant heat flux, $w/m^2$ |
| $c_p$ | specific heat, $kJ/(kg \cdot K)$ |
| $h_{sl}$ | latent heat, $kJ/kg$ |
| $f$ | mass fraction |

**Greek symbols**

| | |
|---|---|
| $\varepsilon_0$ | initial porosity |
| $\varphi$ | volume fraction |
| $\theta$ | dimensionless temperature |
| $k$ | thermal conductivity, $w/(m \cdot K)$ |
| $\alpha_P$ | thermal diffusivity $k_l/(\rho_P c_P)$, $m^2/s$ |

**Subscript**

| | |
|---|---|
| g | gas |
| s | solid |
| l | liquid |
| p | particle |
| pl | liquid phase of particle |
| ps | solid phase of particle |
| 0 | heating surface |